\input{aipcheck}

\documentclass[
    ,final            
  ]
  {aipproc}

\layoutstyle{6x9}

\newcommand{\md}{{\rm d}}
\newcommand{\tr}{{\rm tr}}

\begin{document}

\title{Quantum gravity and cosmological observations}


\author{Martin Bojowald}{
  address={Institute for Gravitational Physics and Geometry, The Pennsylvania State University, 104 Davey Lab, University Park, PA 16801, USA}
}

\begin{abstract}
  Quantum gravity places entirely new challenges on the formulation of
  a consistent theory as well as on an extraction of potentially
  observable effects. Quantum corrections due to the gravitational
  field are commonly expected to be tiny because of the smallness of
  the Planck length. However, a consistent formulation now shows that
  key features of quantum gravity imply magnification effects on
  correction terms which are especially important in cosmology with
  its long stretches of evolution. After a review of the salient
  features of recent canonical quantizations of gravity and their
  implications for the quantum structure of space-time a new example
  for potentially observable effects is given.
\end{abstract}

\maketitle

\section{Gravity}

The gravitational field is the only known fundamental force not yet
quantized despite of more than six decades of research.  Difficulties
arise due to two key properties: Although gravity is the dominant
player on cosmic scales, it is weak in usual regimes of particle
physics. Strong quantum gravity effects, possibly accessible to
observations, thus require large gravitational fields which are
realized only in exotic situations such as the very early universe or
black holes. But then, the classical field grows without bound,
implying space-time singularities.  Secondly, gravity is conceptually
very different from other interactions due to the equivalence
principle: gravity is a manifestation of space-time geometry. The full
space-time metric $g_{\mu\nu}$ is the physical object to be quantized,
not perturbations $h_{\mu\nu}$ on a background space-time such as
Minkowski space $\eta_{\mu\nu}$.

Thus, any quantization of gravity able to describe these phenomena
faithfully must be non-perturbative and background independent for
most applications. A new framework is required which does not
refer to causality or vacuum states and other concepts which are
available only once a metric has already been specified. One has to
make use of the quantum structure of space and time itself.

Although mathematically involved, this is now available in broad form
due to research in the past 15 years. It is not yet uniquely
formulated, but several characteristic properties have been revealed.
All this is essential for non-singular quantum space-times, making
sense even at the big bang, but also for quantum corrections in strong
field regimes as they might be observable as remnants of the very
early universe.  The stage is thus provided for the first potentially
observable effects of quantum gravity.

\section{Possible effects?}

There is a dimensional argument which usually is taken as proof that
quantum gravity effects are tiny, too tiny to be observable anytime
soon. Given Planck's constant $\hbar$ and Newton's constant $G$, one
can (in units where the speed of light is $c=1$) define the Planck
length $\ell_{\rm P}=\sqrt{G\hbar}\approx 10^{-33}{\rm cm}$. Its value
is tiny compared to any scales we can probe directly, or equivalently
the Planck mass $M_{\rm P}=\sqrt{\hbar/G}\approx 10^{19}{\rm GeV}$ is
huge compared to the mass of any elementary object.  Only negligible
correction terms are then expected from dimensionless combinations of
available length scales, e.g.\ of the order $\ell_{\rm P}H\approx
10^{-60}$ in cosmology with the current Hubble length
$H^{-1}=a/\dot{a}$.  Indeed, correction terms in low energy effective
actions, obtained by perturbative approaches on a background
space-time, give only such negligible terms in equations of motion
\cite{CMBShortDist}.

However, for quantum gravity the low energy effective action is too
special unless one is interested only in scattering of gravitons.  A
low energy effective action is obtained by an expansion around the
vacuum state of quantum field theory on a background. The concept of a
vacuum itself changes in background independent approaches since a
vacuum state, defined e.g.\ as the unique Poincar\'e invariant state
of a quantum representation, refers to symmetries of a background
space-time. (Or, as the ground state of the Hamiltonian a vacuum is
uniquely defined only for time-independence, which requires a
symmetry.) Moreover, the gravitational Hamiltonian as it arises from
the action is always unbounded from below and thus lacks a ground
state in the usual sense.

There is an additional expectation from quantum gravity, namely that
space has a discrete structure on very small scales. One can think of
this structure as an irregular lattice whose typical plaquette size
$p$ is close to $\ell_{\rm P}^2$. But unlike the Planck length, this is
a geometrical parameter or field specifying the quantum gravity state
and can thus be dynamical. This parameter brings in crucial
information from quantum gravity, unlike $\ell_{\rm P}$ which is
determined simply by parameters of quantum mechanics and classical
gravity.

In such a situation, there are three length parameters: the
macroscopic scale $H^{-1}$, the fundamental scale $p$ and the
dimensional Planck length $\ell_{\rm P}$.  Any dimensional argument
must then fail since dimensionless combinations of three length
parameters can have any value depending on which geometric mean
$l_1^{x_1}l_2^{x_2}l_3^{x_3}$ with $x_1+x_2+x_3=0$ is relevant in a
given situation. In other words, a large dimensionless parameter
exists such as the number $N$ of lattice sites which may enter and
magnify correction terms.  The precise form of corrections can then
only be determined by detailed calculations taking into account the
discrete structure of space. (See also \cite{Planck} for a critique of
dimensional arguments in quantum gravity.)

\section{Space-time structure}

A quantum theory for space-time structure is required, which is not
unlike atomic pictures in condensed matter physics.
From the two basic properties of general relativity we obtain two
important lessons for the formulation of quantum gravity:
\begin{itemize}
\item Do not use a split $g_{\mu\nu}=\eta_{\mu\nu}+h_{\mu\nu}$ of the
  space-time metric as a background $\eta_{\mu\nu}$ with a
  perturbation $h_{\mu\nu}$, as this does not allow one to describe a
  lattice structure. One rather has, at a fundamental level, metrics
  which are distributional and supported only on lattice links and
  vertices.
\item Do not use low energy effective theory but suitable
  generalizations. Low energy theory does not apply due to the
  unboundedness of the Hamiltonian, and it would even be unclear what
  ``low energy'' refers to given that there are no observer
  independent concepts of energy in general relativity.
\end{itemize}
All these techniques are now available in the framework of loop
quantum gravity \cite{Rov,ALRev,ThomasRev}. In particular,
cosmological effects are calculable with recent progress. The six main
ingredients in this general scheme are described in what follows.

\subsection{1. New variables}

A canonical formulation of general relativity was originally developed
in ADM variables \cite{ADM} given by the spatial metric $q_{ab}$ and
momenta related to extrinsic curvature $K_{ab}$ (or $\dot{q}_{ab}$).
This refers to a chosen time coordinate, but the theory is independent
of that choice if certain constraints are satisfied. These constraints
are equivalent to Einstein's equation and implement general
covariance.  However, when trying to quantize a theory whose dynamical
variable is the metric, it is difficult to turn tensor components such
as $q_{ab}$ itself into operators. In generally covariant systems one
has to take into account arbitrary, non-linear changes of coordinates
$x'(x)$.  A tensor such as $q_{ab}$ then transforms as $q_{a'b'}=
(\partial x^a/\partial x'{}^{a'}) (\partial x^b/\partial x'{}^{b'})
q_{ab}$ which leads to coordinate dependent factors. A quantization of
gravity needs to represent the field $q_{ab}$ and its momenta as
operators on a Hilbert space, but the definition must be independent
of spatial coordinates which are not defined on the Hilbert space.
This is the key difference to quantum field theories defined on, say,
Minkowski space. This background space-time allows only Poincar\'e
transformations $x'(x)$ as coordinate changes, which are linear. The
coordinate change of tensors is then a simple linear transformation by
spatially constant matrices, which can easily be extended to
operators.

This problem, as it turned out, can be circumvented by using a  new
set of variables \cite{AshVar,AshVarReell}:
  $q_{ab}$ is replaced by a co-triad $e_a^i$ (three co-vector
  fields such that $e_a^ie_b^i=q_{ab}$), which then defines the
densitized triad $E^a_i=|\det e_b^j|e^a_i$. Similarly, one defines
$K_a^i:=e^b_iK_{ab}$ and then the connection
  $A_a^i=\Gamma_a^i- K_a^i$ with
 $\Gamma_a^i= -\epsilon^{ijk}e^b_j (\partial_{[a}e_{b]}^k+
 {\textstyle\frac{1}{2}} e_k^ce_a^l\partial_{[c}e_{b]}^l)$.

\subsection{2. Basic objects}

These are variables as in non-Abelian gauge theories, the
Ashtekar--Barbero connection $A_a^i$ canonically conjugate to
``electric fields'' $E^a_i$: $\{A_a^i(x),E^b_j(y)\}=8\pi G
\delta_a^b\delta_j^i\delta(x,y)$. The gauge group here is the group of
triad rotations $E^a_i\mapsto R^j_iE^a_j$, $R\in{\rm SO}(3)$, which
leave the metric unchanged.  Moreover, for any curve $e$ and surface
$S$ in space we define holonomies and fluxes \cite{LoopRep}
\begin{equation}
 h_e(A)={\cal P}\exp\int_e A_a^i\tau_i\dot{e}^a\md t\quad,\quad
 F_S(E)=\int_S\md^2y n_aE^a_i \tau_i
\end{equation}
with the tangent vector $\dot{e}^a$ of $e$, the co-normal $n_a$ of $S$
and Pauli matrices $\tau_i$.  These quantities have many advantages.
(i) They do not have any indices and are thus scalar quantities. No
complicated tensor transformations under non-linear coordinate
transformations arise. (ii) A smearing is automatically included,
resulting in well-defined Poisson brackets to become commutators, free
of any delta functions. (iii) No reference is made to any metric other
than that determined by $E^a_i$, and background independence of the
classical theory is thereby respected.

\subsection{3. Representation}

One can thus construct a quantum representation of these smeared basic
fields and then impose the necessary constraints as operators.  A
convenient way to do this is the connection representation, where
holonomies are used as multiplication operators ``creating'' spin
networks \cite{RS:Spinnet} $T_{g,j,C}(A)=\prod_{v\in g} C_v\cdot
\prod_{e\in g} \rho_{j_e}(h_e(A))$ as an orthonormal basis of
$h_e$-dependent functions, where $g$ is an oriented graph with labels
$j$ as SU(2) representations $\rho_j$ on edges, and $C$ as gauge
invariant contraction matrices in vertices.

\subsection{4. Discrete geometry} 

Fluxes are conjugate to holonomies and thus become derivative
operators
\begin{equation}
 \hat{F}_S^i f_g = -8\pi iG\hbar \int_S {\rm d}^2y n_a
\frac{\delta}{\delta A_a^i(y)} f_{g}(h(A))
 =-8\pi i\ell_{\rm P}^2\sum_{e\in g}\int_S {\rm d}^2yn_a
\frac{\delta (h_e)^A_B}{\delta
    A_a^i(y)}\frac{\partial f_{g}(h)}{\partial(h_e)^A_B}
\end{equation}
when acting on a state $f_g$ as a linear combination of spin networks.
With $\int_S\md^2yn_a\delta h_e/\delta A_a^i(y)=\frac{1}{2}\tau_i
\int_S\md^2y\int_e\md t n_a(y) \dot{e}^a \delta(e(t),y) h_e$ we have
non-zero contributions only if $S$ intersects edges of $g$, and
contributions are determined by su(2) derivatives
$J_i=\tr(\tau_ih\partial/\partial h)$ acting on holonomies. As
invariant derivatives on a compact group (identical to angular
momentum operators), a discrete spectrum of fluxes and thus discrete
spatial geometry results.  This also extends to other geometrical
operators such as the area operator, obtained as a quantization of
$A(S)=\int_S \md^2y \sqrt{E^a_in_a E^b_in_b}$, \cite{AreaVol,Area}
\[
 \hat{A}(S)f_{g,j} = 4\pi\ell_{\rm P}^2\sum_{p\in S\cap g}
 \sqrt{j_p(j_p+1)} f_{g,j}
\]
(assuming no intersections in vertices of $g$). Similarly the volume
operator has a discrete spectrum receiving contributions only from
vertices in a region \cite{AreaVol,Vol2}.

The representation used so far is not only an explicitly known one,
but is in fact, to a large degree, unique using the covariance under
spatial diffeomorphisms \cite{LOST,WeylRep}.  As it happens
sometimes also in quantum field theories on a background, symmetry
reduces the freedom in choosing a representation. The large symmetry
group of diffeomorphisms present due to background independence
selects a unique representation. This differs from usual Fock spaces
for particle excitations, but Fock states can be reproduced as
distributional states.  There is thus a tight kinematical setting,
although quantization ambiguities such as factor ordering will as
usually arise for composite operators, in particular the constraints,
specifying the dynamics.

\vspace{-5mm}

\subsection{Example: Cubic lattice}

\vspace{-2mm}

Labels $j$ appearing on edges of graphs determine the geometry through
flux values of the densitized triad. They are half-integer and thus
imply a minimum non-zero flux given by $4\pi\ell_{\rm P}^2$. The
states are all that is present in the quantum theory to determine
geometry. The lattice thus {\em is space,} and no continuum limit is
to be taken as one would do it in lattice gauge theories.  From the
labels we obtain lattice fluxes $p_{v,I}=F_{S_{v,I}}=8\pi\ell_{\rm
  P}^2 j_{v,I}$ depending on the direction $I$ of an edge and its
starting vertex $v$. This is the state dependent scale of geometry
introduced before as the additional length scale provided by discrete
quantum gravity. This scale can be inhomogeneous if labels differ much
for different edges, and it is dynamical (as well as, possibly, the
lattice structure such as the number of vertices). States, in this
way, determine which physical scales are relevant. Recent developments
from different directions within loop quantum gravity have now
converged to such structures \cite{APSII,AQGI,InhomLattice}.

\subsection{5. Gravitational dynamics}

Dynamics of space-time is determined by the Hamiltonian constraint
which, when solved, is supposed to show which special superpositions
of lattices are allowed for generally covariant states.  This
constraint is implemented through lattice Hamiltonians which change
the labels and possibly the graph when acting on a state. Thanks to
spatial discreteness, those operators are well-defined even with
matter contributions: there are no UV divergences \cite{QSDI,QSDV}.
As even classical expressions are complicated non-polynomial functions
of the basic fields, the operators are only barely tractable. Although
they have been defined rigorously, they have quantization ambiguities
(such as factor ordering and several other choices) and do not easily
reveal interesting solutions.  But they do display characteristic
properties which follow from spatial discreteness and are common to
all available constructions. They can be tested with suitable
approximation schemes, which currently include symmetries
\cite{SymmRed,LivRev} or perturbations
\cite{InhomLattice,QuantCorrPert}.

More in detail, the classical expression of the constraint functional
(to be imposed for all spatial functions $N(x)$) is
\begin{equation} \label{Ham}
 H[N] = \frac{1}{16\pi G} \int_{\Sigma} \mathrm{d}^3x N
 \left(\epsilon_{ijk}F_{ab}^i\frac{E^a_jE^b_k}{\sqrt{|\det
E|}}  -
4
 (A_a^i-\Gamma_a^i)(A_b^j-\Gamma_b^j)
\frac{E^{[a}_iE^{b]}_j}{\sqrt{|\det E|}}
 \right)
\end{equation}
which requires an inverse determinant of $E^a_i$. This is not
available immediately due to the fact that fluxes, and also the volume
operator, have discrete spectra containing zero and thus no inverse
operators. But one can use the identity \cite{QSDI}
\begin{equation} \label{ident}
 \left\{A_a^i,\int{\sqrt{|\det E|}}\mathrm{d}^3x\right\}= 2\pi G
 \epsilon^{ijk}\epsilon_{abc} \frac{E^b_jE^c_k}{{\sqrt{|\det E|}}}
\end{equation}
to obtain a well-defined quantization through a Poisson bracket, which
then becomes a commutator of holonomies and volume.  For the curvature
components $F_{ab}^i$ we use $s_1^as_2^b F_{ab}^i\tau_i=
\Delta^{-1}(h_{\alpha}-1) +O(\Delta)$ and write it in terms of a
holonomy $h_{\alpha}$ around a square loop of coordinate size $\Delta$
and with tangent vectors $s_1^a$ and $s_2^a$ at $v$ \cite{RS:Ham}.
Finally, an extrinsic curvature operator for $A_a^i-\Gamma_a^i$ in
(\ref{Ham}) can be derived as a double commutator.

\subsection{6. Effective theory}

In any quantum field theory, especially one with a highly complicated
Hamiltonian, progress can be made only with suitable approximations to
compute physical effects. One of the main tools in particle physics is
the low energy effective action which allows powerful applications for
instance in perturbation theory.  The lattice Hamiltonians we have
here are different from quantum field theory Hamiltonians on a
background, and conceptually also from lattice gauge theory. For
instance, they are unbounded from below already classically. No ground
state is available to expand around as done in low energy effective
actions.  It is thus necessary to generalize effective theory which
has been accomplished \cite{EffAc,Karpacz}. Effective equations, in
general terms, are obtained from an analysis of the coupled dynamics
of $n$-point functions.  (In quantum mechanics, these are spread and
deformations of a wave packet back-reacting on the peak position.)
Effective dynamics is given by the expectation value of the
Hamiltonian in suitable semiclassical states, with a precise
specification depending on the regime to be analyzed.

Applied to lattice Hamiltonians, we can already draw one important
conclusion: local coefficients of the effective Hamiltonian appear as
functions of $\ell_{\rm P}^2/p_{I,v}\gg \ell_{\rm P}^2H^2$ (after,
e.g., using (\ref{ident}) with local lattice building blocks).
Quantum effects are thus much larger than they would be in low energy
effective theory due to the discrete structure \cite{InhomEvolve}.

\subsection{Summary of Quantum effects}

Typical properties of effective Hamiltonians as they are also known
from effective actions are \cite{Karpacz}: (i) Factors such as the
quantized inverse metric determinant give rise to modified small-scale
behavior of coefficients (possibly related to boundedness of
classically diverging curvature expressions near singularities
\cite{InvScale,Sing}).  (ii) Replacing local curvature and connection
components by holonomies along extended loops implies non-locality or,
when Taylor expanded in an effective Hamiltonian, higher order spatial
derivatives.  (iii) The coupling of $n$-point functions in general
effective equations implies, as usually, new quantum degrees of
freedom (related to higher time derivatives).

Properties (i) and (ii) are typical holonomy effects of the loop
quantization which was forced upon us by background independence while
(iii) is a genuine quantum effect. Both (ii) and (iii) correspond to
higher curvature terms, while (i) corrects geometrical factors purely
from quantum geometry.

\section{Application: How big is the typical quantum scale?}

Lattice states as solutions to the Hamiltonian constraint are
difficult to find even numerically. But orders of magnitude of
corrections can be estimated based on two different roles played by
the fundamental scale $p$.  First, $p$ determines the number of
lattice sites within a typical macroscopic scale which we take to be
$H^{-1}$: for larger $p$, the lattice is coarser and discreteness
corrections arise.  {\em Continuum physics} requires $p\ll H^{-2}$.
Secondly, the size of $p$ signals quantum effects since it is
proportional to the quantum number of a state. For smaller $p$, a
lower ``excitation level'' is realized and thus one has larger quantum
effects. {\em Semiclassical physics} requires $p\gg\ell_{\rm P}^2$.

These are two opposite requirements, leaving an allowed range
$\ell_{\rm P}^2\ll p\ll H^{-2}=\frac{3}{8\pi G\rho}$ where we took the
Hubble length as the macroscopic scale relevant for cosmology, and
computed it in terms of energy density $\rho$ using the Friedmann
equation.  This range is wide in the late universe, where quantum
corrections can be almost arbitrarily small.  But it is more narrow in
the early universe and during inflation where the typical energy scale
implies $1\gg \ell_{\rm P}^2/p\gg 10^{-6}$.

Direct observations of effects from $\ell_{\rm P}^2/p\approx 10^{-6}$
would not be observable soon, but magnifications can occur if they add
up during long cosmic evolution.  This is in fact realized
\cite{InhomEvolve} as it follows from cosmological perturbation
equations derived from the effective Hamiltonian \cite{HamPerturb}:
One of the metric modes, $\Phi$, satisfies
 $\ddot{\Phi}+(1+\nu)\dot{\Phi}/\eta
+ \epsilon\Phi/\eta^2=0$
on large scales as a differential equation in conformal time $\eta$
and with $\nu$ being related to the matter equation of state $w$. The
quantum correction $\epsilon$ changes the behavior crucially compared
to the classical situation where $\epsilon=0$.  Solutions for constant
$\nu$ are $\Phi(\eta)=\eta^{\lambda}$ with $\lambda=-\frac{\nu}{2}\pm
\frac{1}{2} \sqrt{\nu^2-4\epsilon}$. For $\epsilon=0$, one mode is
constant, one decays, which is known as conservation of curvature
perturbations.  Loop quantum gravity implies $\epsilon<0$, such that
the constant mode becomes slightly growing and curvature perturbations
are not exactly preserved.

Since $\epsilon$ has a definite sign, which is a robust property in
loop quantum gravity independently of quantization ambiguities
\cite{Ambig,ICGC}, small corrections indeed add up during evolution.
During inflation, conformal time for the largest visible modes changes
by a factor $e^{-60}$. Thus, a factor $e^{-60\epsilon}\approx
1+10^2|\epsilon|$ results, magnifying the correction to at least
$10^{-4}$ for $\epsilon$, as estimated above, of the size $10^{-6}$.

\section{Summary}

Effects from the basic scale of quantum gravity in phenomenological
equations can thus be computed with recent advances, and they reveal
sometimes surprising implications. Classical modes (or also gravitons)
are not fundamental but collective excitations out of the microscopic
discrete state. Basic excitations are rather the scale parameters
$p_{v,I}$ which, when excited inhomogeneously, can give rise to the
classical fields or, in some approximation, gravitons on large scales.

Systematic calculations are now possible, mainly due to advances in
the understanding of effective theory, and can be applied to different
regimes. While characteristic effects are robust under quantization
ambiguities, ambiguity parameters might become relevant for possible
observations with more precise data.  Cosmological observations thus
have a good chance of revealing quantum gravity effects in the near
future and to guide further constructions to complete the theory.

\begin{theacknowledgments}
  The author thanks Hugo Morales-T\'ecotl and Luis Urrutia for an
  invitation to give a plenary talk at the VIth Latin American
  Symposium on High Energy Physics (Puerto Vallarta, Mexico, Nov.\
  2006), on which this contribution is based. Results reported here
  were obtained through work funded by NSF grant PHY0554771.
\end{theacknowledgments}


\end{document}